**Sunspot Count Periodicities in Different Zurich Sunspot Group Classes since 1986**

A. Kilcik[1] · A. Ozguc[2] · V.B. Yurchyshyn[3] · J.P. Rozelot[4]

[1]Faculty of Science, Department of Space Science and Technologies, Akdeniz University, 07058 Antalya, Turkey
e-mail: alikilcik@akdeniz.edu.tr
[2]Kandilli Observatory and Earthquake Research Institute, Bogazici University, 34684 Istanbul, Turkey
[3]Big Bear Solar Observatory, Big Bear City, CA 92314, USA
[4]Nice University, OCA- Bd de l'OBSERVATOIRE, CS 34229, 6304 NICE CEDEX 4, France

**Abstract**
In this study, we used two methods to investigate the periodic behavior of sunspot counts in four categories for the time period January 1986-October 2013. These categories include the counts from simple (A and B), medium (C), large (D, E, and F), and final (H) sunspot groups. We used: i) the Multi-taper Method with red noise approximation, and ii) the Morlet wavelet transform for periodicity analysis. Our main findings are: (1) the solar rotation periodicity of about 25 to 37 days, which is of obvious significance, is found in all groups with at least a 95% significance level; (2) the periodic behavior of a cycle is strongly related to its amplitude and group distribution during the cycle; (3) the appearance of periods follow the amplitude of the investigated solar cycles, (4) meaningful periods do not appear during the minimum phases of the investigated cycles.
We would like to underline that the cyclic behavior of all categories is not completely the same; there are some differences between these groups. This result can provide a clue for the better understanding of solar cycles.

**Introduction**
Sunspots are the dark magnetic structures continuously observed on the solar surface since 1600s. Thus, they are the oldest and the most commonly used indicator of solar activity. Analysis of the observed sunspot numbers (SSNs) and other solar activity indicators such as sunspot area (SSA), solar flare index (FI), total solar irradiance (TSI), 10.7 cm solar radio flux (F10.7), etc, indicated that the solar activity displays periodic variations from days to thousands of years.
The existence of the 11-year solar cycle and the 27-day solar rotation periodicities are well known. The interval between 27 days and 11 years is called the 'mid-range' periodicities (Bai, 2003). Searches for additional possible periodicities, other than 27 days and 11 years in solar activity indicators have been of interest for a long time. Many researchers have investigated these periodicities using various solar activity parameters and found periods such as 450-512 days (1.2-1.4 years), 280-364 days (0.8-1.0 years), 210-240 days (0.6-0.7 years), 150-170 days (0.4-0.5 years), 120-130 days, 110-115 days, 73-78 days, 62-68 days, 51-58 days, 41-47 days, and 25-37 days (Rieger et al., 1984; Lean and Brueckner, 1989; Pap, Bouwer, and Tobiska, 1990; Bai and Sturrock, 1991; Bouwer, 1992; Prabhakaran Nayar *et al.*, 2002; Krivova and Solanki, 2002; Ozguc, Atac, and Rybak, 2003, 2004; Kane, 2003, 2005; Bai,

2003; Obridko and Shelting, 2007; Chowdhury, Khan, and Ray, 2009; Kilcik *et al.,* 2010; Chowdhury and Dwivedi, 2011; Scafetta and Willson, 2013 and references therein).

Nevertheless, there are still disagreements between the periods determined by different studies probably due to problems in the analyzed data, the methods used, investigated time intervals, *etc*. These differences are possibly related to the different origin of various solar activity indicators (Bouwer, 1992). Detection of a periodicity and its source may provide a clue for a better understanding of solar activity.

It is well known that the modified Zurich classification is based on three components (McIntosh, 1990). The first component is the sunspot group class (A, B, C, D, E, F, and H), which describes the size of a sunspot group and the distribution of the penumbrae. The second component describes the largest spot in a group, and the third one is the compactness of sunspots in the intermediate part of a group (for more detail see McIntosh, 1990). Active regions were divided into two groups, small and large groups for the first time by Kilcik *et al.,* (2011a). They found that these groups behave differently along a solar cycle. Here we focus on the first component of modified Zurich classification (similar to their approximation) and separate sunspot groups in four categories which are: simple (A and B), medium (C), large (D, E, and F) and final (H) groups. The separation is based on the size and time evolution of the groups. We investigated the periodic behavior of sunspot counts in these groups, and found that they exhibit differences during the investigated time interval.

In Section 2, we describe the methodology and the data analysis. In Section 3, we present the results. In Section 4, we draw conclusions and briefly discuss their implications.

**Data and Methods**
The data used in this study were downloaded from the National Geophysical Data Center (NGDC: ftp://ftp.ngdc.noaa.gov/STP/SOLAR\_DATA); they include the group classification, sunspot count, group extension, SSA, name of observatory, observation quality, magnetic classification, NOAA identification number, and time information for each recorded group during the observed day. The data are collected by the United States Air Force$/$Mount Wilson Observatory (USAF/MWL). This database also includes measurements from the Learmonth Solar Observatory, the Holloman Solar Observatory, and the San Vito Solar Observatory. We used the Holloman station data as the principal data source and gaps were filled with records from one of the other stations listed above, so that a nearly continuous time series was produced.

As mentioned in the Introduction, sunspot groups have been divided into four categories depending on their size and time evolution: The first category is formed by simple groups, which include A and B classes. They do not have penumbrae, and are in the early stage of their evolution. The second category is formed by medium groups that include only the C class. They have penumbrae at one end of the group, and are in the middle of their group evolution. The third category is formed by large groups that include D, E and F groups. They have penumbrae at both sides of the group, and are quite developed compared to the previous ones. The fourth category is formed by final groups, which include only the H class, and they are at the end of the sunspot time evolution. To investigate the periodic behavior of these categories, sunspot count data were used for the time period January 1986-October 2013, which covers Cycles 22 (1986-1996), 23 (1996-2008) and almost half of Cycle 24. This time interval contains a total of 10288 days.

To analyze the periodic behavior of each category we used the Multi-taper Method (MTM) and Morlet wavelet analysis with a red noise approximation. The MTM was applied to obtain the exact values of the periods, but it does not give any information about the operation time (temporal variations of the periods) of a periodic behavior. To remove this uncertainty and check the existence of such periods, Morlet wavelet analysis was also applied.

**Multi-taper Method**
The MTM was first developed by Thomson (1982) for reducing the variance of spectral estimates by using a set of tapers and to recover the lost information, while still maintaining an acceptable bias. The method uses orthogonal windows (or tapers) to obtain approximately independent estimates of the power spectrum and then combines them to yield an estimate. This estimate exhibits more degrees of freedom and allows an easier quantification of the bias and variance trade--offs, compared to the conventional Fourier analysis. It has been previously used successfully for the analysis of different time series with different time scales and resolutions such as climatic data (Ghil *et al.*, 2002; Wilson *et al.*, 2007; Kilcik, Ozguc, and Rozelot, 2010; Marullo, Artale, and Santoleri, 2011; Fang *et al.*, 2012; Escudier, Mignot, and Swingedouw, 2013), solar data (Prestes *et al.*, 2006; Kilcik *et al.*, 2010; Mufti and Shah, 2011) *etc.* The MTM used as a harmonic analysis permits us to detect low--amplitude harmonic oscillations in relatively short time series with a high degree of statistical significance. It also allows us to reject larger amplitude harmonics if the F test (variance test) fails. This feature is an important advantage of the MTM over other classical methods for which the error bars scale with the peak amplitudes (Jenkins and Watts, 1968; Ghil *et al.*, 2002).

In this study we used three sinusoidal tapers, and the frequency range was taken from 0.0014 to 0.04 (*i.e.* 25-712 days). The selection of this time interval is based on two main criteria: i) the smallest detectable period must include at least the time length of a solar rotation, and ii) the selection of 712 days as an upper limit will allow us resolve the solar rotation period inside the largest period. The significance test was carried out assuming that the noise has a red spectrum, since the data have a larger power density at lower frequencies and a smaller power density at higher frequencies (thus longer periods have more noise than smaller ones). The method was thus applied in the selected interval of frequencies with a half-band width resolution of two and a tapering of three. The detected signals were kept when a 95% confidence level was reached.

**Morlet Wavelet Analysis**
The Morlet wavelet method is a powerful tool for analyzing localized power variations within a time series. This method has been extensively applied in solar physics *e.g.* Torrence and Compo, 1998; Krivova and Solanki, 2002; Lundstedt *et al.*, 2006; Li *et al.*, 2011; Deng *et al.*, 2013; Choudhary *et al.*, 2014 and references therein). Contrary to the classical Fourier analysis that decomposes a signal into different sine and cosine functions which are not bounded in time, the wavelet transform uses wavelets characterized by scale (frequency) and time localization. It can be employed to analyze time series that contain non-stationary powers at many different frequencies. According to Templeton (2004) *this analysis method has great utility in several areas of astronomy and astrophysics, since many objects have varying periods, or have no fixed period at all and instead show transient periodicities or quasi-periodicities*. We used the Morlet wavelet analysis to determine possible temporal variations in the periodicities that were found in the sunspot counts from the MTM analysis. The Morlet wavelet is a complex sine wave, localized within a Gaussian window (Morlet *et al.,* 1982). Its frequency domain representation is a single symmetric Gaussian peak, and its localization is very accurate. The use of such a wavelet has the advantage of incorporating a wave with a clear period, as well as finite in extent. The sunspot count data have been analyzed using the standard Interactive Data Language (IDL) package for Morlet wavelet analysis, and the scalograms were obtained to study both the presence and evolution of the periodicities. This analysis was carried out using the full daily data. The smallest scale selected as 16 days.

**Results**
In this work we focus on the temporal variations of the sunspot counts in groups over a time interval that includes three solar cycles. This time period begins on January 1, 1986 and ends on October 31, 2013 (a total of 10288 days). The MTM spectral analysis and Morlet wavelet analysis methods were used to find the periodicities and their operation time for each group daily data during the studied time interval, respectively. The MTM analysis was performed using a red noise approximation and with 95% confidence level. To study the temporal evolution of the entire range of obtained periods in sunspot counts, the Morlet wavelet transform was also applied using the red noise approximation with 90% confidence since the wavelet transform method gives detailed information on the time localization of each periodicity. Figures 1, 2, 3, and 4 show the results of the MTM analysis and the wavelet scalograms of all the sunspot counts of the investigated groups during the entire time interval. After a close inspection of these figures, there are several interesting points that we want to emphasize: 1) The solar rotation periodicity appears in all groups and cycles (Cycle 22 and 23 and the ascending branch of Cycle 24) for the investigated time period. It is interesting to note that the smallest solar rotation period obtained for the medium groups was 29 days, while this periodicity appears as 25-26 days in all other groups. 2) The appearance of periods follows the solar cyclic behavior. No meaningful period appears during the minimum phases (around 1986, 1996 and 2008 years in the wavelet scalogram) of the investigated cycles. On the contrary, all periods appear as dominant during the cycle maxima. 3) The appearance of the periods follows the amplitude of the investigated solar cycles; more periods appear in the cycle with the largest maximum (Cycle 22), there are less in the one with the lower maximum (Cycle 23), and even less in the lowest one (the ascending branch of Cycle 24). 4) The number of present periods shows a remarkable difference depending on the type of sunspot group and the amplitude of the solar cycle. 5) Similar to the solar rotation periodicity, about 55-day periodicities exist in all groups, while all other periodicities have a group preference (see Table1).

Figure 1 shows the MTM and wavelet analysis results for simple sunspot groups. All the periodicities that appear in MTM power spectrum (Figure 1a) exist in the wavelet scalogram (Figure 1b) of Cycles 22 and 23 as well, while no periodicity greater than 76 days appears during the ascending branch of Cycle 24.

Figure 2 shows the same results for the sunspot counts in medium groups. Contrary to Figure 1, a relatively large periodicity (about 250 days) appears during the ascending branch of Solar Cycle 24, while the small periods almost disappear. In addition, a relatively large number of periodicities exists (compared to small groups) during Cycle 23. Contrary to other groups, medium sunspot groups shows the largest significant periodicity of 348 days.

Figure 3 shows the results for large sunspot groups. The interesting point in this figure is that the appearance of periods in Cycles 22 and 23 are similar, while only one periodicity (about 26 days) appears during the ascending phase of Cycle 24.

Finally, in Figure 4 we present the results for the final sunspot groups. Here, all three cycles are similar regarding the appearance of the periods. There are small gaps during the minima, and all periods appear in all cycles.

In Table 1, we present a list of significant periods and their appearance in different groups. A close look at the table shows us three significant points: that we would like to emphasize; first, some periods (about 51-59, 25-37 days) exist in all groups; second, 129-168 days (the Rieger periods), only appear in small and final sunspot groups; third, a 41 days period appears only in final sunspot groups.

**Discussion and Conclusions**
The existence of different periods in selected sunspot groups may be indicative of the true solar cyclicity. Its understanding may contribute to predict the solar activity more accurately and, correspondingly, the geomagnetic activity. We discuss our findings in the following sections.

**Consistency with Earlier Results**
We have found the following periods: 25-37, 41, 48, 51-59, 65, 74-79, 100, 112, 129, 168, 213, 248, 315-348, 420, and 480 days, with at least 95% confidence. The solar rotation period (about 27 days) is present in all the investigated groups. Periodicities close to 27 days may be attributed to the effect of solar rotation, and they may appear longer or shorter due to the slight variations in the active region evolutions. On the other hand the results of the spectral analysis can be spurious, since the solar time series are not fully stationary (see Velasco, Mendoza, and Valdes-Galicia, 2008). Kilcik *et al.* (2010) analyzed solar FI data for Cycles 21, 22, and 23 using the MTM and the Morlet wavelet analysis; they found periods of 25-30, 62, 115, 137, 248 and 410 days with at least 90% confidence for the maximum phase of Cycle 23. By applying the wavelet transform method to the daily relative SSN over Cycles 10-23, Yin *et al.* (2007) found that a period of about 27 days existed in almost all solar cycles, and even during minima.
Chowdhury, Khan, and Ray (2009) have investigated the periodicities of SSAs. They reported periods of 24-45, 69-95, 113-133, 160-187, 245-321, 348-406 days and about 1.3 years. Lou

*et al.* (2003) have found periods of 33.53 (±0.52), 38.71 (±0.55), 42.16 (± 0.78), 63.73 (± 2.34), 67.69 (± 2.18), 98.18 (±3.25), 122.19 (± 4.88), 156.77 (± 10.89), and 259.48 (± 24.23) days by applying the Fourier power spectral analysis to X-ray flares exceeding M5 class during the maximum phase of Cycle 23 (1999-2003). Chowdhury and Dwivedi (2011) investigated the periodicities of SSNs and the coronal index by using the wavelet power spectrum technique for the time span from May 1996 to December 2008. They found a solar rotation period of about 27 days, plus periods of 40-60, 50-70, 60-80, 100, 110-130, 150-160 days and about 1.3 years in different time intervals.

Recently, Scafetta *et al.* (2013) analyzed the TSI data by adopting a multi-scale dynamical spectral analysis technique from 2003.15 to 2013.16 (descending phase of Cycle 23 and ascending phase of Cycle 24). These authors found periods of 0.070 years (25.6 days), 0.095 years (34.7 days), 0.20 years (73 days), 0.25 years (91 days), 0.30-0.34 years (109.6-124.2 days), 0.39 years (142.4 days) and 0.75-0.85 years (273.9-310.5 days). As a result, all the periodic variations found in this article were obtained by using different solar activity indicators and different methods. All these earlier studies are a confirmation of our findings. As shown in Table 1, the solar rotation (about 27 days) and and a period of 51-59 days appear in all groups. Thus, we conclude that these two periodicities do not have a group preference.

**Group Preference of Sunspot Count Periodicities**
All periodicities except two of them (about 27 and 51-59 days) have a group preference. A periocity of about 55 days has been found by different authors using different solar activity indicators (Kilcik *et al.,* 2010; Chowdhury and Dwivedi, 2011, and references therein). A periodicity of 323 days, close to our 315 days, has been found by Lean and Brucker (1989) using from all solar activity indicators (the sunspot blocking function, Zurich sunspot numbers, F10.7, and $Ca_{II}$ K plage index). These authors concluded that this periodicity can have a real solar origin. Our findings confirm their results. A period of about 100 days exists in all groups except large ones, one of 315-348 days also appears in all groups but simple ones, and one of about 76 days is also present in all groups but final ones. About 1.3-year and 1.1-year periods do not appear in sunspot count of medium groups. It is interesting to note that a period of 213 days appears only in medium groups, while one of 41 days is only present in final groups. In of our analysis, we have found the Reiger periods of 129-168 days. As shown in Table 1 these periodicities do not appear in medium and large group data. The 112-day periodicity appears only in sunspot counts of simple and final groups. A 100-day periodicity appears in all groups but large ones, while an 85-day periodicity appears in both simple and final groups. Periods of about 74-79 and 65 days do not appear in final groups. Another interesting result is that medium groups show more short periods (<80 days) as compared to other categories. As discussed above, all these periodicities were found earlier using different data sets and methods. In this article, we analyze the group preference of periodic variations of sunspot counts, for the first time. As a result, we speculate that the cyclic behavior of sunspot counts in different sunspot groups is different, and may be related to sunspot evolution.

**Solar Activity and Periodicity Relationships**

We start this section pointing out the results of Yin *et al.* (2007) concerning SSNs for Cycles 10-23. It is clearly seen in their Figure 2 that the appearance of periods found were related to the cycle phases; rare periods appeared during the minimum phase (see valley in their Figure 2), while most of the periods were present during the maximum phase. Our results confirm their findings in the following way; during the beginning of our analyzed cycles (Cycle 22, 23, and 24) the wavelet scalograms look almost "clean", while they are more complicated during the maximum phases.

Another interesting point is the appearance of periods. They follow the cycle amplitude; strong/intense cycles have a larger number of periodic variations, while weak ones have smaller. Kilcik *et al.* (2011a) separated sunspot groups in two classes as large and small. They compared the monthly large and small group numbers, and found that the number of large groups was comparable during Cycles 22 and 23, while the number of small groups had a strong decrease during Cycle 23. These authors were able to explain the unusual behavior of Cycle 23 as the results of the great number of large groups present during this cycle (Kilcik *et al.,* 2011a, 2011b). Later, Lefevre and Clette (2012) analyzed each Zurich class from A to F, except H, separately. They concluded that there was a strong deficit of small groups during Solar Cycle 23 compared to Cycle 22. It is interesting to note that the appearance of periods for the selected categories follow this variation. As shown in our wavelet scalograms, the number of observed periodic behaviors is very small in sunspot counts of simple and medium groups during Solar Cycles 23 and 24, while it is comparable in large and final groups during Cycle 23. Thus, we speculate that most of the periodic variations during Cycle 23 come mainly from the two latter groups. During Cycle 24, the situation is different. The variations mainly come from the sunspot counts of final groups; all other categories present almost "clean" wavelet scalograms. Thus, we conclude that the periodic behavior during a cycle is strongly related to the cycle amplitude and the observed group distribution during that cycle.

Kilcik *et al.* (2011a) separated sunspot groups in two classes as large and small. They found that large groups reach their maximum number about two years later than small ones. Recently, Kilcik and Ozguc (2014) investigated the origin of a double peak solar cycle maximum using the sunspot group classification. They concluded that a double peak maximum could originate in the different behavior of large and small groups. In this article, we have found that different groups have different cycle behaviors as discussed in Section 4.2. The 231-day and 41-day periodicities appear only in medium and final groups while 100-day periodicity appears in all groups except large ones. Therefore, the behavior of periodicities can be used to distinguish between the first and second peaks of a solar cycle. Thus, we conclude that periodicity investigations based on separating the sunspot groups can provide more accurate information about a cycle and a clue for a better understanding of solar cycles.

Finally, the main findings of our study are as follows:

1. Periodic behaviors during a cycle are strongly related to the cycle amplitude and the observed group distribution.

2. Similar to the solar rotation periodicity, an about 55-day periodicity exists in all sunspot groups while all other periodicities have a group preference.

3. The appearance of periods are related to the amplitude of investigated solar cycles, such that periodic behaviors are more frequent during the cycle with the largest maximum (Cycle 22), less frequent during the one with a lower maximum (Cycle 23), and even less frequent in the lowest one (Cycle 24).

4. The appearance of periods is related to the solar cycle evolution in such way that no meaningful periods appear during the minimum phases of the investigated cycles; while, on the contrary, all periods are dominant around the cycle maxima.

**Acknowledgement**
The authors thank to anonymous referee for his/her useful comments and suggestions. The wavelet software was provided by C. Torrence and G. Compo, and is available at http://paos.colorado.edu/research/wavelets. We acknowledge usage of sunspot data from the National Geophysical Data Center. This study was supported by the Scientific Research Projects Coordination Unit of Akdeniz University through Project of 2012.01.0115.008.

**References**
Bai, T.: 2003b, *Astrophys. J.* **591**, 406. doi: 10.1086/375295
Bai, T., and Sturrock, P.A.: 1991, *Nature*, **350**, 141. doi: 10.1038/350141a0
Bouwer, S.D.: 1992, *Solar Phys.* **142**, 365. doi: 10.1007/BF00151460
Chowdhury, P., Khan, M., Ray, P.C.: 2009, *Mon. Not. Roy. Astron. Soc.* **392**, 1159. doi: 10.1111/j.1365-2966.2008.14117.x
Chowdhury, P. and Dwivedi, B.N.: 2011, *Solar Phys.* **270**, 365. doi: 10.1007/s11207-011-9738-1
Choudhary, D.P., Lawrence, J.K., Norris, M., Cadavid, A.C.: 2014, *Solar Phys.* **289**, 649. doi: 10.1007/s11207-013-0392-7
Deng, L.H., Qu, Z.Q., Yan, X.L., Wang, K.R.: 2013, *Res. in Astron. Astrophys.* **13**, 104. doi: 10.1088/1674-4527/13/1/011
Escudier, R., Mignot, J., Swingedouw, D.: 2013, *Clim. Dyn.* **40**, 619. doi 10.1007/s00382-012-1402-4
Fang, K., Gou, X., Chen F., Liu, C., Davi, N., Li, J., Zhao, Z., Li, Y.: 2012, *Global and Planetary Change* **80**, 190. doi:10.1016/j.gloplacha.2011.10.009
Ghil, M., Allen, M.R., Dettinger, M.D., Ide, K., Kondrashov, D., Mann, M.E., *et al.*: 2002, *Rev. Geophys.* **40**, 1. doi: 10.1029/2000RG000092
Jenkins, G.M., Watts, D.G.: 1969, *Holden-Day Series in Time Series Analysis: Spectral analysis and its applications*, Holden-Day, London.
Kane, R.P.: 2003, *J. Atmos. Solar-Terr. Phys.* **65**, 979. doi: 10.1016/S1364-6826(03)00117-2
Kane, R.P.: 2005, *Solar Phys.* **227**, 155, doi: 10.1007/s11207-005-1110-x
Kilcik, A., Ozguc, A., Rozelot, J.P.: 2010, *J. Atmos. Solar-Terr. Phys.* **72**, 1379. doi: 10.1016/j.jastp.2010.10.002
Kilcik, A., Ozguc, A., Rozelot, J.P., Atac,T.: 2010, *Solar Phys.* **264**, 255. doi: 10.1007/s11207-010-9567-7
Kilcik, A., Yurchyshyn, V.B., Abramenko, V., Goode, P., Ozguc, A., Rozelot, J.P., Cao, W.: 2011, *Astrophys. J.* **731**, 30. doi: 10.1088/0004-637X/731/1/30 .
Kilcik, A., Yurchyshyn, V.B., Abramenko, V., Goode, P., Gopalswamy, N., Ozguc, A., Rozelot, J.P.: 2011, *Astrophys. J.* **727**, 44. doi: 10.1088/0004-637X/727/1/44
Kilcik, A., and Ozguc, A.: 2014, *Solar Phys.* **289,** 1379, doi: 10.1007/s11207-013-0407-4
Krivova, N. A. and Solanki, S. K.: 2002, *Asron. Astrophys.* **394**, 701-706, doi: 10.1051/0004-6361:20021063
Lean, J.L., and Brueckner, G.E.: 1989, *Astrophys. J.* **337**, 568. doi: 10.1086/167124
Lefevre, L., Clette, F.A.: 2011, *Astron. Astrophys.* **536**, L11. doi: 10.1051/0004-6361/201118034

Li, K.J., Shi, X.J., Liang, H.F., Zhan, L.S., Xie, J.L., Feng, W.: 2011, *Astrophys. J.* **739**, 49. doi: 10.1088/0004-637X/730/1/49

Lou, Y.Q., Wang, Y.M., Fan, Z., Wang, S., Wang, J.X.: 2003, *Mon. Not. Roy. Astron. Soc.* **345**, 809. doi: 10.1046/j.1365-8711.2003.06993.x

Lundstedt H., Liszka, L., Lundin, R. , Muscheler, R.: 2006, *Ann. Geophys.* **24**, 769. doi: 10.5194/angeo-24-769-2006

Marullo, S., Artale, V., Santoleri, R.: 2011, *J. Climate* **24**, 4385. doi: 10.1175/2011JCLI3884.1

McIntosh, P.S.: 1990, *Solar Phys.* **125**, 251. doi: 10.1007/BF00158405

Morlet, J., Arehs, G., Forgeau, I., Giard, D.: 1982, *Geophysics* **47**, 203. doi: 10.1190/1.1441328

Mufti, S., Shah G.N.: 2011, *J. Atmos. Solar-Terr. Phys.* **73**, 1607. doi:10.1016/j.jastp.2010.12.012

Obridko, V.N. and Shelting, B.D.: 2007, *Adv. Space Res.* **40**, 1006. doi: 10.1016/j.asr.2007.04.105

Ozguc, A., Atac, T., Rybak, J.: 2003, *Solar Phys.* **214**, 375. doi: 10.1023/A:1024225802080

Ozguc, A., Atac, T., Rybak, J.: 2004, *Solar Phys.* **223**, 287. doi: 10.1007/s11207-004-7304-9

Pap, J., Bouwer, S.D., Tobiska, W.K.: 1990, *Solar Phys.* **129**, 165. doi: 10.1007/BF00154372

Prabhakaran Nayar, S.R., Radhika, V.N., Revathy, K., Ramadas,V.: 2002, *Solar Phys.* **208**, 359. doi: 10.1023/A:1020565831926

Prestes, A., Rigozo, N.R. Echer, E. Vieira, L.E.A.: 2006, *J. Atmos. Solar-Terr. Phys.* **68**, 182. doi: 10.1016/j.jastp.2005.10.010

Rieger, E., Kanbach, G., Reppin, C., Share, G.H., Forrest, D.J., Chupp, E.L.: 1984, *Nature* **312**, 623. doi: 10.1038/312623a0

Scafetta, N., and Willson, R.C. 2013, *Pattern Recognition in Physics* **1**, 123. doi: 10.5194/prp-1-123-2013

Templeton, M.: 2004, *J. Am. Assoc. Var. Star Obs.* **32**, 41

Thomson, D.J.: 1982, *Proc. IEEE* **70**, 1055

Torrence, C., Compo, G.P.: 1998, *Bull. Am. Meteorol. Soc.* **79**, 61. doi: 10.1175/1520-0477(1998)079<0061:APGTWA>2.0.CO;2

Velasco, V.M., Mendoza, B., Valdes-Galicia, J.F.: 2008, In: Caballero, R., D'Olivo, J.C., Medina-Tanco, G., Nellen, L., Sánchez, F.A., Valdés-Galicia, J.F. (eds.) Proc. *30th Int. Cosmic Ray Conf.* **1**, 553

Wilson, R., Wiles, G., Rosanne D., Zweck, C.: 2007, *Clim. Dyn.* **28**, 425. doi: 10.1007/s00382-006-0194-9

Yin, Z.Q., Han, Y.B., Ma, L.H., Le, G.M., Han, Y.G.: 2007, *Chin. J. Astron. Astrophys.* **7**, 823. doi: 10.1088/1009-9271/7/6/10

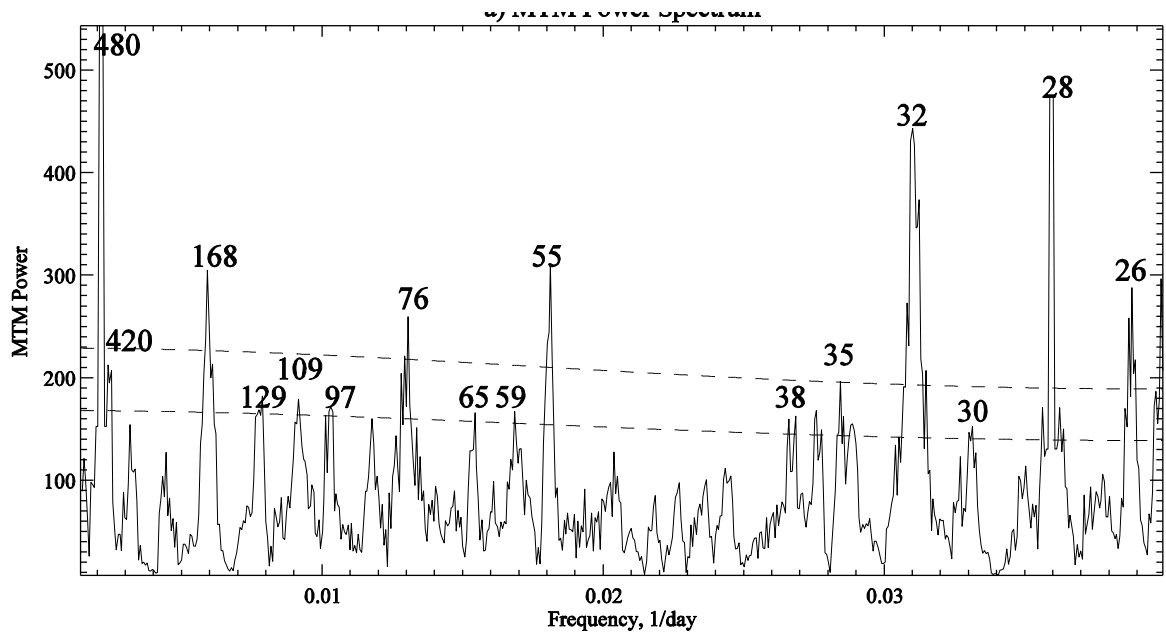

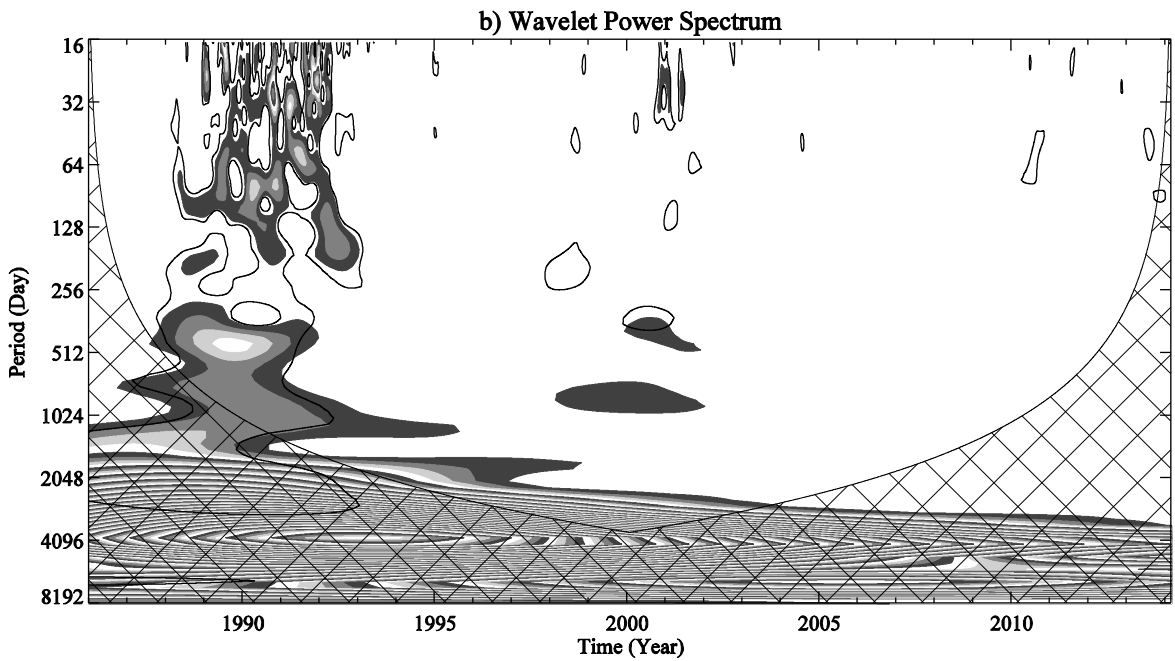

**Figure 1**. (a) and (b) analysis of cycle periods using the MTM and the Morlet wavelet method for the daily sunspot counts in small groups, respectively. The data extend from January 1986 to October 2013. The black contours in the wavelet scalogram (b) indicate a 90% confidence level and the hatched area below the thin black line is the cone of influence (COI). Numbers close to the peaks in (a) show the value of periods as day.

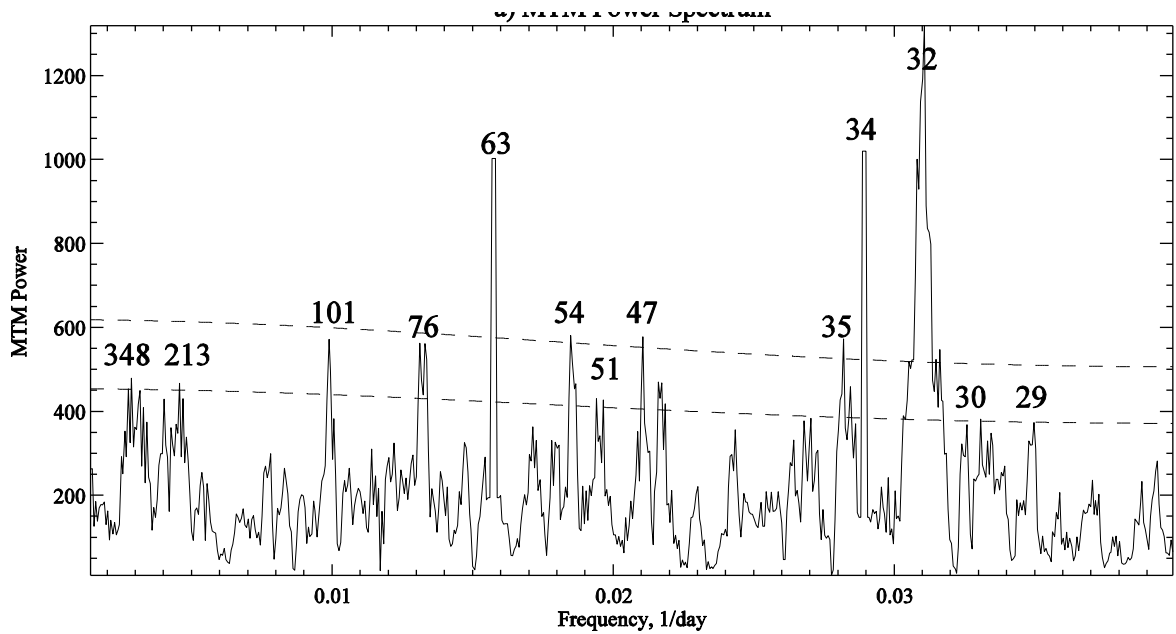

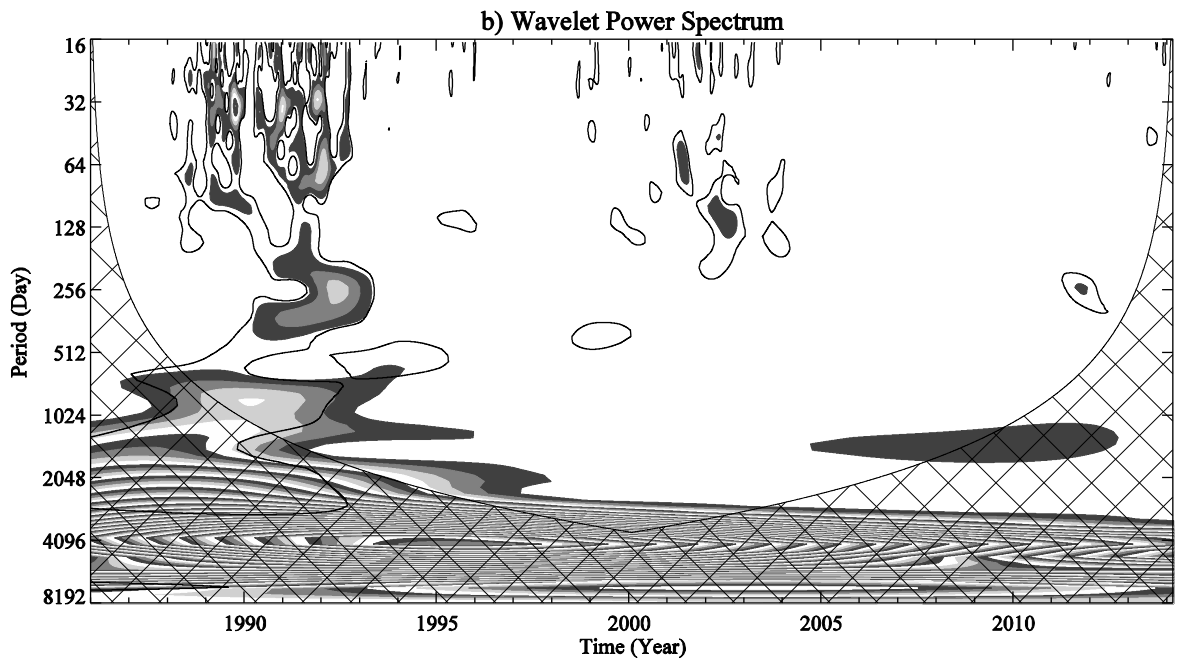

**Figure 2.** (a) and (b) analysis of cycle periods using the MTM and the Morlet wavelet method for the daily sunspot counts in medium groups, respectively. The data extend from January 1986 to October 2013. The black contours in the wavelet scalogram (b) indicate a 90% confidence level and the hatched area below the thin black line is the cone of influence (COI). Numbers close to the peaks in (a) show the value of periods as day.

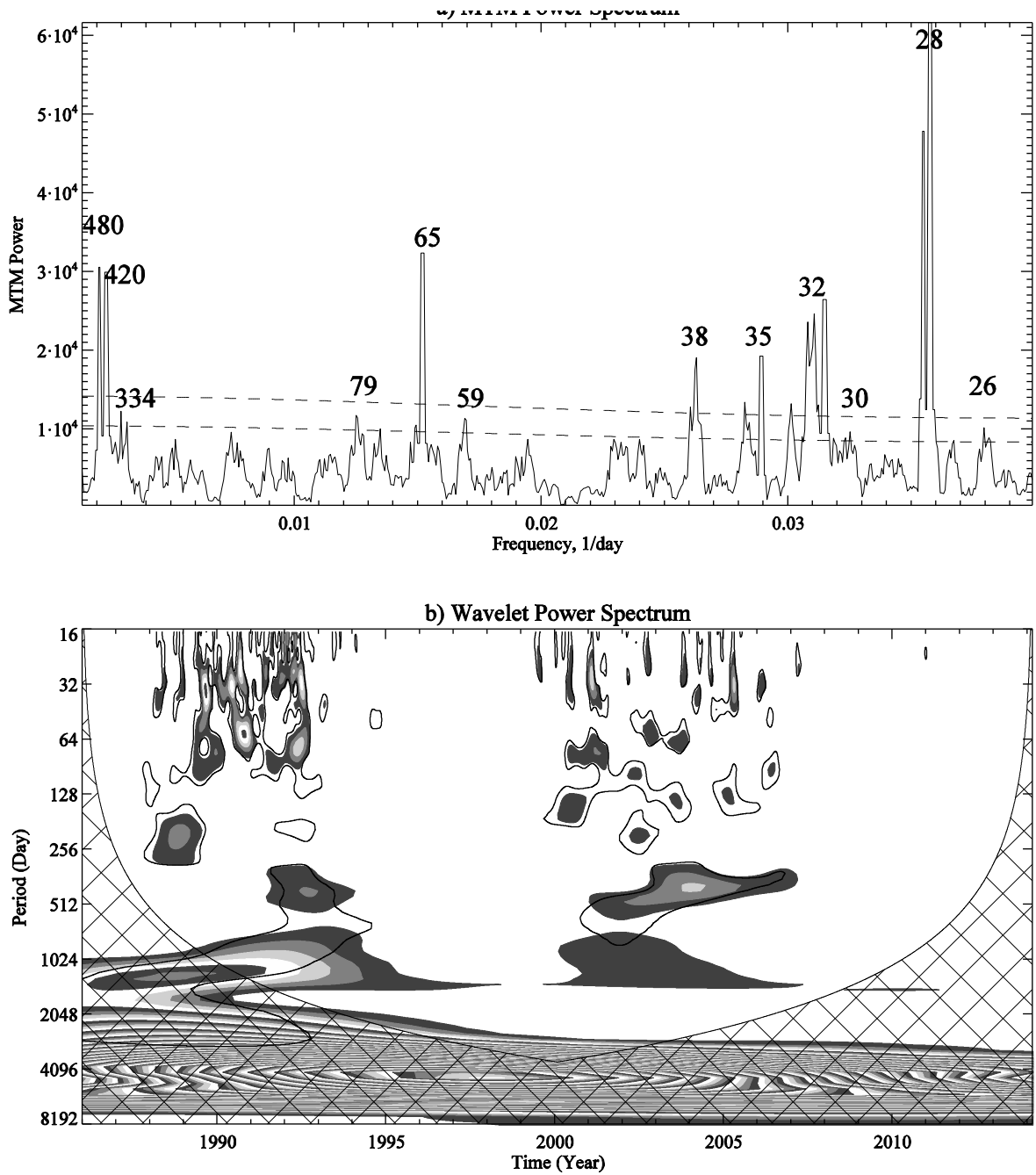

**Figure 3.** (a) and (b) analysis of cycle periods using the MTM and the Morlet wavelet method for the daily sunspot counts in large groups, respectively. The data extend from January 1986 to October 2013. The black contours in the wavelet scalogram (b) indicate a 90% confidence level and the hatched area below the thin black line is the cone of influence (COI). Numbers close to the peaks in (a) show the value of periods as day.

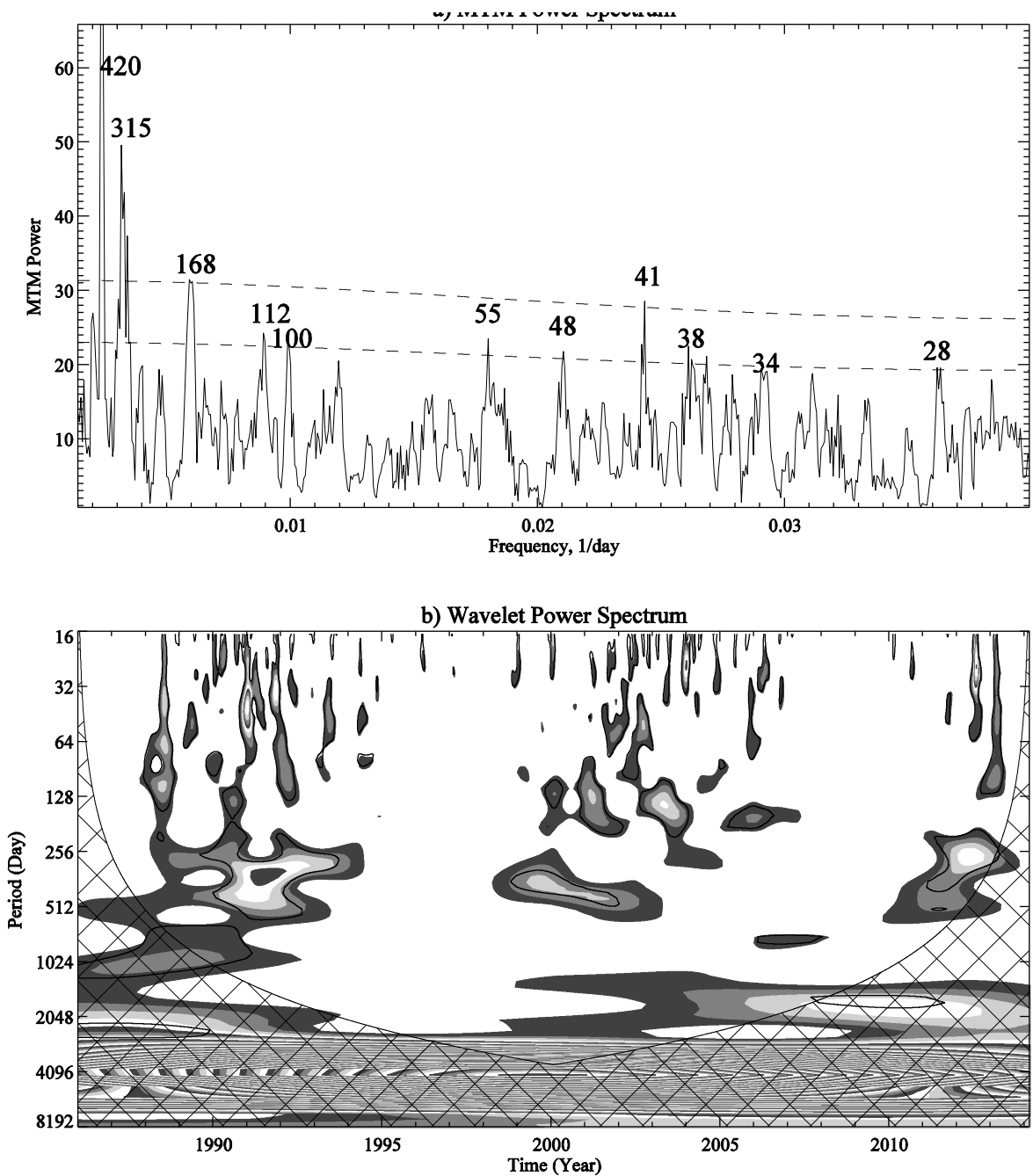

**Figure 4.** (a) and (b) analysis of cycle periods using the MTM and the Morlet wavelet method for the daily sunspot counts in final groups, respectively. The data extend from January 1986 to October 2013. The black contours in the wavelet scalogram (b) indicate a 90% confidence level and the hatched area below the thin black line is the cone of influence (COI). Numbers close to the peaks in (a) show the value of periods as day.

**Table 1.** Periods obtained for different groups using the MTM. The first column corresponds to the obtained periods and the rest of the columns show the presence of these periods in all groups by means of their significance levels. The numbers between parentheses show the closest values in the other groups.

| Period (Day) | Simple Groups | Medium Groups | Large Groups | Final Groups |
|---|---|---|---|---|
| 480 | +>99 | - | + >99 | + >95 (496) |
| 420 | + >95 | - | + >99 | + >99 |
| 315-348 | - | + >95 (348) | + >95 (334) | >99 (315) |
| 213 | - | + >95 | - | - |
| 168 | + >99 | - | - | + >99 |
| 129 | + >95 | - | - | - |
| 112 | + >95 (109) | - | - | + >95 |
| 100 | + >95 (97) | + >95 | - | + >95 |
| 79 - 74 | + >99 | + >95 | + >95 | - |
| 65 | + >95 | + >99 (63) | + >99 | - |
| 51-59 | + >95 (55&59) | + >95 (54&51) | + >95 (59) | +>95 (55) |
| 48 | - | + >99 | - | + >95 |
| 41 | - | - | - | + >99 |
| 25-37 | + >95 | + >95 | + >95 | + >95 |